\documentclass[mathleft
]{an}
\usepackage{graphicx}
\usepackage{times}
\begin{document}

% The following seven commands are intended for editorial usage and should be ignored by
% the author(s).
%\Pagespan{789}{}% Document's page range. 
\Pagespan{1}{}% Document's page range. 
% If second parameter is left empty, the last page is computed automatically.
%\Yearpublication{2006}%
%\Yearsubmission{2005}%
%\Month{11}%   
%\Volume{999}%  
%\Issue{88}% 
\Yearpublication{ }%
\Yearsubmission{2010}%
\Month{ }%   
\Volume{ }%  
\Issue{ }% 
% \DOI{This.is/not.aDOI}% 

\title{Optical variability of the ultraluminous X-ray source NGC 1313 X-2}

\author{D. Impiombato\inst{1}, L. Zampieri\inst{1}\fnmsep\thanks{Corresponding author: \email{luca.zampieri@oapd.inaf.it}\newline},
R. Falomo\inst{1}, F. Gris\'e\inst{2} \and R. Soria\inst{3}
%Example 
%for footnote, note the usage of the \texttt{fnmsep}
%command as separator between institute number and footnote mark} 
}
\titlerunning{Optical Variability of NGC 1313 X-2}
\authorrunning{D. Impiombato, L. Zampieri, R. Falomo, F. Gris\'e \& R. Soria}
\institute{
INAF-Osservatorio Astronomico di Padova, Padova I-35122, Italy
\and 
Department of Physics and Astronomy, University of Iowa, Van Allen Hall, Iowa City, IA 52242, USA
\and 
MSSL, University College London, Holmbury St Mary, Surrey RH5 6NT, UK
}

\received{ }
\accepted{ }
\publonline{ }


\abstract{%
We analyzed the longest phase-connected photometric dataset available for NGC 1313 X-2,
looking for the $\sim$6 day modulation reported by Liu et al. (\cite{l09}). The folded $B$ band
light curve shows a 6 days periodicity with a significance slightly larger than $3\sigma$.
The low statistical significance of this modulation, along with the lack of detection in 
the $V$ band, make its identification uncertain.
}

\maketitle

\begin{table}
\label{tab1}
% \centering
% \begin{minipage}{140mm}
\caption{Log of the VLT+FORS1 and HST+WFPC2 photometric observations of NGC 1313 X-2}
\begin{tabular}{ccccc}\hline
 Obs. &  Date  &  MJD  &  Exposure  &  Instr.  \\
      &        &       &  (s)       &          \\
 \hline
 1  &    2007-10-21   & 54394.362587   & 242$\times$2 &  FORS1  \\
 2  &    2007-11-15   & 54419.220472   & 242$\times$2 &  FORS1  \\
 3  &    2007-11-15   & 54419.277624   & 242$\times$2 &  FORS1  \\
 4  &    2007-11-16   & 54420.255469   & 242$\times$2 &  FORS1  \\
 5  &    2007-12-06   & 54440.075805   & 242$\times$2 &  FORS1  \\
 6  &    2007-12-10   & 54444.173203   & 242$\times$2 &  FORS1  \\
 7  &    2007-12-14   & 54448.121299   & 242$\times$2 &  FORS1  \\
 8  &    2008-01-31   & 54496.073081   & 242$\times$2 &  FORS1  \\
 9  &    2008-03-02   & 54527.059338   & 242$\times$2 &  FORS1  \\
 10 &    2008-03-05   & 54530.053776   & 242$\times$2 &  FORS1  \\
 11 &    2008-03-08   & 54533.056052   & 242$\times$2 &  FORS1  \\
 12 &    2008-05-21   & 54607.911122   & 500$\times$2 &  WFPC2  \\
 13 &    2008-05-22   & 54608.043761   & 500$\times$2 &  WFPC2  \\
 14 &    2008-05-23   & 54609.042372   & 500$\times$2 &  WFPC2  \\
 15 &    2008-05-24   & 54610.040983   & 500$\times$2 &  WFPC2  \\
 16 &    2008-05-25   & 54611.038900   & 500$\times$2 &  WFPC2  \\
 17 &    2008-05-26   & 54612.104178   & 500$\times$2 &  WFPC2  \\
 18 &    2008-05-27   & 54613.179178   & 500$\times$2 &  WFPC2  \\
 19 &    2008-05-28   & 54614.178483   & 500$\times$2 &  WFPC2  \\
 20 &    2008-05-29   & 54615.177789   & 500$\times$2 &  WFPC2  \\
 21 &    2008-05-30   & 54616.097233   & 500$\times$2 &  WFPC2  \\
 22 &    2008-05-31   & 54617.109039   & 500$\times$2 &  WFPC2  \\
 23 &    2008-06-01   & 54618.692372   & 500$\times$2 &  WFPC2  \\
 24 &    2008-06-02   & 54619.173622   & 500$\times$2 &  WFPC2  \\
 25 &    2008-06-03   & 54620.105567   & 500$\times$2 &  WFPC2  \\
 26 &    2008-06-04   & 54621.171539   & 500$\times$2 &  WFPC2  \\
 27 &    2008-06-05   & 54622.104178   & 500$\times$2 &  WFPC2  \\
 28 &    2008-06-06   & 54623.102789   & 500$\times$2 &  WFPC2  \\
 29 &    2008-06-07   & 54624.034733   & 500$\times$2 &  WFPC2  \\
 30 &    2008-06-08   & 54625.100706   & 500$\times$2 &  WFPC2  \\
 31 &    2008-06-09   & 54626.611817   & 500$\times$2 &  WFPC2  \\
\hline                                                  
\end{tabular}
\end{table}

\section{Introduction}

NGC 1313 X-2 is located in the outskirts of the barred spiral galaxy NGC 1313 at a distance 
of 3.7–-4.27 Mpc (Tully \cite{t88}; M\'endez et al. \cite{m02}; Rizzi et al. \cite{r07}).
Its observed X-ray luminosity varies between a few $\times 10^{39}$ erg/s and $\sim 10^{40}$
erg/s in the 0.3–-10 keV band (Feng \& Kaaret \cite{f06}). The source has been 
extensively studied in the X-ray and optical bands (e.g. Mucciarelli et al. \cite{m07}; Gris\'e et 
al. \cite{g08}). It belongs to a handful of ultraluminous X-ray sources (ULXs) clearly associated to stellar optical counterparts
(e.g. Liu et al. \cite{l04}; Kaaret et al. \cite{k04}; Mucciarelli et al. \cite{m05}; Soria et al. \cite{s05}). 
These sources appear almost ubiquitously hosted in young stellar environments (e.g. Pakull et al. \cite{p06},
Ramsey et al \cite{r06}, Liu et al. \cite{l07}) and have properties consistent with those of young, 
massive stars. However, some ULXs appear to be associated to older stellar populations and 
one possible later type stellar counterpart is now known (Feng \& Kaaret \cite{f08}; 
Roberts, Levan \& Goad \cite{r08}), although its spectral classification may be affected by
significant galactic and extra-galactic reddening (Gris\'e et al. \cite{g06}).
%although there are possible counter-examples (e.g. Feng \& Kaaret \cite{g08}; Roberts, Levan \& Goad \cite{r08}). 
In the case of NGC 1313 X-2, a single optical counterpart has been identified 
through a chain of efforts (Zampieri et al. \cite{z04}; Mucciarelli et  al. \cite{m05,m07}; Pakull et al. 
\cite{p06}; Liu et al. \cite{l07}; Gris\'e et al. \cite{g08}). This star has an extinction-corrected absolute magnitude 
$M_B\sim -4.5$ mag and colors $(B-V)_0\sim -0.15$ mag and $(V-I)_0\sim -0.16$ mag
(Mucciarelli et al. \cite{m07}; Gris\'e et al. \cite{g08}), consistent with a B spectral type.

\begin{figure*}
\includegraphics[width=170mm]{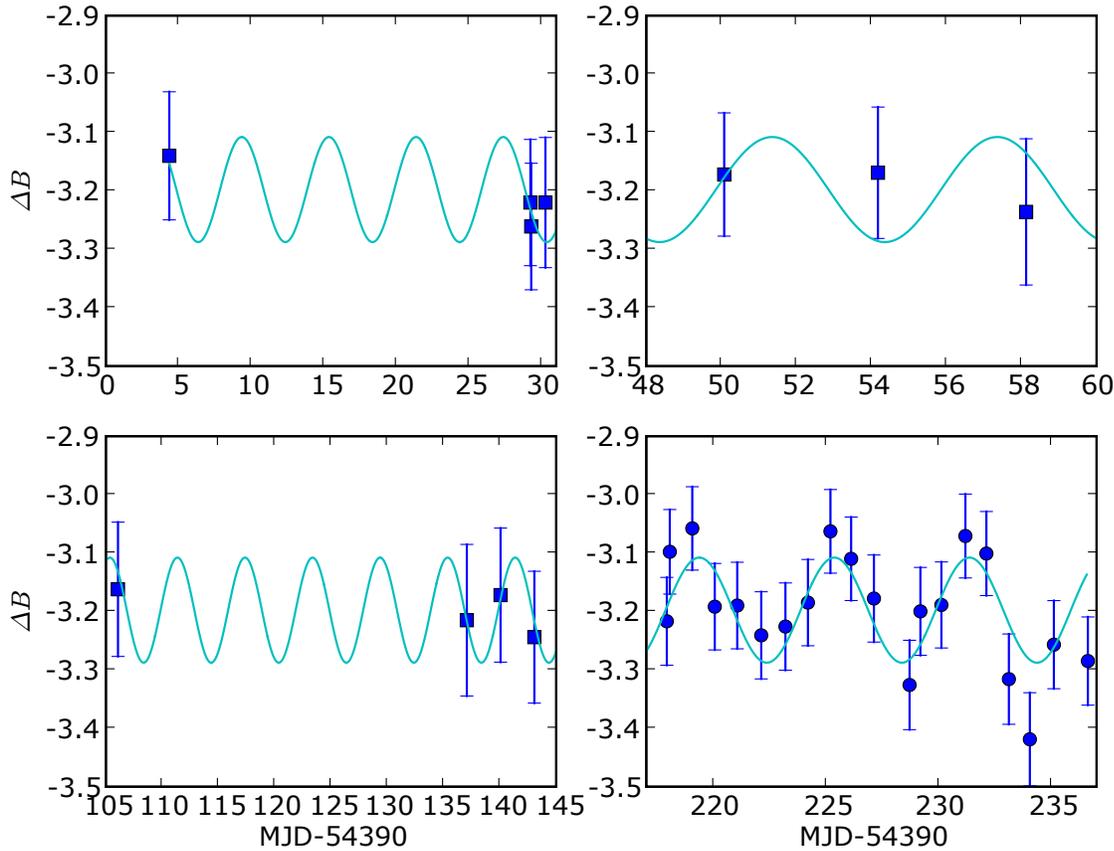}
\caption{Joint VLT+HST light curve of NGC 1313 X-2 in the $B$ band. The squares are the 
VLT+FORS1 data, while the circles represent the HST+WFPC2 observations. The magnitudes 
are the difference with those of a reference field star. The data cover a period of $\sim 7.5$ months. 
The solid ({\it cyan}) line is the best fitting sinusoid with a period  $P=6.0$ days.}
\label{fig1}
\end{figure*}

Liu et al. (\cite{l09}) found a possible periodicity of 6.12$\pm$ 0.16 days in the $B$ band 
light curve of the optical counterpart of NGC 1313 X-2, that was interpreted as the orbital 
period of the binary system. Three cycles were detected in the $B$ band, while no modulation 
was found in $V$.  Previous studies carried out on the available HST and VLT observations  
led to negative results (Gris\'e et al. \cite{g08}). More recently, lack  of significant photometric 
variability on a new sequence of VLT observations has been reported by Gris\'e et al. (\cite{g09}).

Here we present a preliminary reanalysis of the joint VLT+FORS1 and HST+WFPC2 photometric observations 
of NGC 1313 X-2 obtained during the years 2007-2008, with the aim of clarifying the statistical
significance of the orbital periodicity identified by Liu et al. (\cite{g09}). Further details
will be reported in a separate paper (Zampieri et al., in preparation).

\section{VLT and HST observations}

NGC 1313 X-2 was observed with VLT+FORS1 between October 2007 and March 2008 (11 epochs; 
Gris\'e et al. \cite{g09}) and with HST+WFPC2 between May and June 2008 (20 epochs; Liu et al. \cite{l09}). 
The quality of the HST images is fair, despite the degradation of the central PC chip.
A log of the observations is reported in Table~1. We re-analyzed the whole dataset 
in a homogeneous way, looking for the $\sim$6 day periodicity reported by Liu et al. (\cite{l09}).

After performing standard image reduction in the IRAF environment, the two exposures taken 
a few minutes apart each night were combined together and cleaned for cosmic rays. 
To accurately photometer the objects, we used AIDA (Astronomical Image Decomposition and 
Analysis; Uslenghi \& Falomo \cite{u08}), an IDL-based package originally designed to perform 
two-dimensional PSF model fitting of quasar images. For the analysis of the WFPC2 exposures, 
we loaded into AIDA the appropriate PSF simulated with Tiny Tim v. 
6.3\footnote{http://www.stsci.edu/software/tinytim/tinytim.html}.

%\begin{figure*}
%\includegraphics[width=100mm]{phase.eps}
%\caption{Reduced chi-square ($\chi^2_r$) versus phase obtained from the best fit with a sinuosoid of the 
%joint VLT+HST dataset ({\it blue} hexagons, {\it top}) and the HST data alone ({\it black} triangles, {\it bottom}).}
%\label{fig2}
%\end{figure*}

\begin{figure*}
\includegraphics[width=100mm]{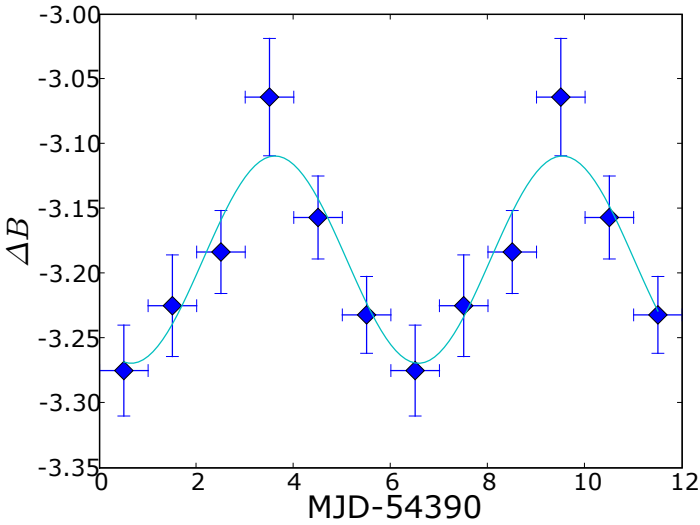}
\caption{Binned light curve (6 bins) of the $B$ band VLT+HST dataset of NGC 1313 X-2, folded 
over the best estimate of the period. The best fitting sinusoid is also shown.}
\label{fig3}
\end{figure*}

Analyzing VLT and HST measurements together requires attention to be paid to
the systematic differences between the two photometric systems.
%including the effects of seeing and of the different bandpasses.
We then first converted the HST instrumental magnitudes 
to the standard $UBVRI$ photometric system using the updated 
transformation equations and coefficients published in Dolphin (\cite{d09}) for the appropriate instrumental 
gain (which is equal to 7 in our case)\footnote{http://purcell.as.arizona.edu/wfpc2$_{-}$calib/}. The color correction term was 
computed adopting the $(B-V)$ color reported in Mucciarelli et al. (\cite{m05}).
In spite of this, residual systematic differences between the two photometric systems might still be present and
affect our measurements. In particular the color correction term is sensitive
to the overall bandpass (telescope plus atmospheric response)
of the instruments. Further attention should be paid to minimize the effects of possible absolute 
calibration uncertainties. We then decided to perform differential photometry of the target with respect to a 
nearby field star (star D in Zampieri et al. \cite{z04}), located on the same chip in both instruments.
The reference star is brighter than the target and has a low root mean square variability ($\sim 0.05$ mag
in the VLT and $\la 0.02$ mag in the HST exposures).
For similar reasons, during all the observations performed with HST+WFPC2, the field was always 
oriented in the same direction and the target and reference star were always located on the 
same position on the central PC chip. 
%At the end we found that the average $B$-band differential 
%magnitude of the VLT and HST are in very good agreement (3.201$\pm$0.038 and 3.196$\pm$0.094, repsectively),
%indicating that any residual systematic uncertainty is probably small in comparison with our present photometric errors.

Figure~\ref{fig1} shows the $B$ band light curve of NGC 1313 X-2 obtained in this way. The data show
clear short term ($\sim 1$ day) variability, likely due to X-ray irradiation. As can be seen from Figure~\ref{fig1}, 
the VLT data have much smaller root mean square variability ($\sim$0.04 mag) than the HST ones 
($\sim$0.09 mag). Thus, there are different levels of optical activity whose origin is unclear
and is under investigation at present (Zampieri et al., in preparation).
%It might be possible that between May and June 2008 the average accretion 
%rate (and hence X-ray irradiation) was slightly larger and/or that the disc emission was partly obscured 
%by material blown off from the inner regions, so that the optical emission underwent more pronounced variations. 
Superimposed on the short term stochastic variability, the HST dataset shows also an approximately sinusoidal 
modulation with a period of 6 days (Figure~\ref{fig1}).

\section{Results}

Following Liu et al. (\cite{l09}), we fitted all the VLT+HST datasets (31 epochs) with a sinusoid:
\begin{equation}
\Delta B = \bar B + A \sin (2\pi (t-t_1)/P + \phi) \, ,
\label{eq1}
\end{equation}
where $A$, $P$ and $\phi$ are the amplitude, period and phase, respectively, $t_1=54390$
is a reference epoch and $\bar B=$ $<\Delta B>=<B_D-B_{target}>=-3.198$ is the average differential magnitude.
The best fitting parameters are: $P=6.01_{-0.01}^{+0.17}$ days, $A=0.09_{-0.02}^{+0.02}$ mag, $\phi={94^0}_{-13^0}^{+13^0}$.
The value of the amplitude and period are in agreement with those reported by Liu et al. (\cite{l09}). The error on
the period is such that $\Delta P (T/P)\la 6$ days, where $T\sim 200$ days is the interval
between the first and last observation (sampling interval). This is not larger than $P$, 
indicating that the two datasets can be phase connected. Although the VLT data 
alone do not show evidence of periodicity, they appear to be consistent with the sinusoidal 
modulation observed in the HST observations. A comparison of the VLT data alone with the 
best fit from the HST data shows that the former are statistically consistent with the HST fit 
(reduced chi-square $\chi^2_r=0.42$).
%Figure~\ref{fig2} shows the reduced chi-square as a function
%of phase obtained from the best fitting sinusoid of the HST data alone (with fixed amplitude 
%and period) and that obtained from the whole VLT+HST dataset. In both cases we recover the same signal.

We reanalyzed also the $V$ band HST+WFPC2 images and found no significant periodic variability.
If the observed modulation is caused mostly by X-ray irradiation, the amplitude in the $V$ band
is expected to be smaller than that in the $B$ band (Patruno \& Zampieri \cite{p10}). A sinusoidal modulation 
with the same period and phase obtained from the fit of the $B$ band data and an amplitude 
$\la 0.04$ mag is consistent, within the errors, with the data (reduced chi-square $\chi^2 \la 1.3$).

The statistical significance of the $B$ band modulation was tested performing a Lomb-Scargle 
periodogram analysis of all the observations. We found that the modulation is significant only 
at the $\la 1.5 \sigma$ level. Binning the light curve in 6 bin intervals and performing an
epoch folding period search, the 6 days modulation is recovered with a significance slightly 
larger than $3 \sigma$ (see Figure~\ref{fig3}). Although the binned light curve suggests that the 
periodicity may be there, the low statistical significance of the $B$ band modulation, along 
with the lack of detection in the $V$ band, make its identification uncertain.
A dedicated photometric monitoring campaign under homogeneous observing conditions to minimize 
systematic uncertainties are needed to confirm it.

\acknowledgements
DI and LZ acknowledge financial support \\ through INAF grant PRIN-2007-26. We thank the referee
for useful comments that helped to improve the paper and the organisers of the ``Ultraluminous X-ray
sources and middle-weight black holes'' workshop for a very interesting and stimulating meeting.

\newpage%%%%%%%%%%%%%%%%%%%%%%%%%%%%%%%%%%%%%%%%%%%%%%%%%%%%%%

\end{document}